\documentclass[aps,prd,twocolumn,superscriptaddress,showpacs]{revtex4}

\usepackage{graphics}

\def\LL{\left\langle}   
\def\RR{\right\rangle}  
\def\LP{\left(}         
\def\RP{\right)}        

\def\BE{\begin{equation}}
\def\EE{\end{equation}}
\def\BEA{\begin{eqnarray}}
\def\EEA{\end{eqnarray}}
\def\EL{\nonumber\\}

\begin{document}


\title{Hybrid configuration content of heavy S-wave mesons}


\author{ Tommy Burch }
\affiliation{Institut f\"ur Theoretische Physik, Universit\"at Regensburg, D-93040 Regensburg, Germany}

\author{ Doug Toussaint }
\affiliation{Department of Physics, University of Arizona, Tucson, AZ 85721, U.S.A.}

\collaboration{MILC Collaboration}
\noaffiliation

\date{\today}

\begin{abstract}
We use the non-relativistic expansion of QCD (NRQCD) on the lattice to study
the lowest hybrid configuration contribution to the ground state of heavy
S-wave mesons. Using lowest-order lattice NRQCD to create the heavy-quark
propagators, we form a basis of ``unperturbed'' S-wave and hybrid states.
We then apply the lowest-order coupling of the quark spin and chromomagnetic
field at an intermediate time slice to create ``mixed'' correlators between
the S-wave and hybrid states. From the resulting amplitudes, we extract the
off-diagonal element of our two-state Hamiltonian. Diagonalizing this
Hamiltonian gives us the admixture of hybrid configuration within the meson
ground state. The present effort represents a continuation of previous work:
the analysis has been extended to include lattices of varying spacings,
source operators having better overlap with the ground states, and
the pseudoscalar (along with the vector) channel.
Results are presented for bottomonium ($\Upsilon$, $\eta_b^{}$)
using three different sets of quenched lattices.
We also show results for charmonium ($J/\psi$, $\eta_c^{}$) from one
lattice set, although we note that the non-relativistic approximation is not
expected to be very good in this case.
\end{abstract}


\maketitle

\section{Introduction}

The existence of valence gluons in bound quark systems, a theoretical
possibility considered for a long time in QCD, continues to elude
confirmation. Allowance for gluonic excitations increases the range of
possible hadronic quantum numbers ($J^{PC}$) beyond those predicted by
constituent quark models. Exotic states should appear (for recent reviews
see Refs.~\cite{Michael:2003ai,Barnes:2003vy}) and, in fact, one such state
($1^{-+}$) has been
observed recently \cite{Ivanov:2001rv}, but the underlying structure of the
state (hybrid meson, four-quark state, meson molecule, etc.) has not been
determined. When considering gluonic constituents, however, the resulting
state need not be exotic; the valence gluons may also combine with the
quarks and antiquarks to form a state which may otherwise be formed without
the gluonic presence. In this non-exotic scenario, a meson (or baryon) state
should consist of a mixture of configurations; not only the case where only
the valence quarks and antiquarks appear, but also those where a gluonic
excitation is present: a hybrid configuration.

As a relevant example -- since this is one of the systems we study in the
present work -- we may look at the vector meson state for bottomonium
($\Upsilon$, $1^{--}$). We may envision the ground state for this system
as a bottom quark and antiquark in a color singlet, a relative S-wave,
and a spin triplet. However, the true ground state should also have a
contribution where the quark and antiquark are in a spin singlet and a color
octet, the spin of the meson and the overall color singlet being ensured by
the gluonic excitation:
\BEA
|\Upsilon\rangle &=& \hspace*{0.27cm} {\cal A}_s |b\bar b\rangle
\hspace*{1.02cm} + \hspace*{0.19cm} {\cal A}_h |b\bar bg\rangle \EL
&=& \cos\theta |1S(1^{--})\rangle + \sin\theta |1H(1^{--})\rangle .
\EEA
It is just this type of (albeit simplified) two-state system we consider in
the present work. We also consider the $0^{-+}$ heavy S-wave meson:
\BE
|\eta_b^{}\rangle = \cos\theta' |1S(0^{-+})\rangle +
\sin\theta' |1H(0^{-+})\rangle .
\EE

Hybrid-quarkonium configuration mixing has been considered before in the
framework of the MIT bag model \cite{Barnes:1979hg,deViron:1984su}
and with the use of an adiabatic potential model \cite{Gerasimov:1998tm}.
We compare results from this on-going lattice work
\cite{Burch:2001tr,Burch:2001nk,Burch:2003zz} with these previous results.

\section{Lattice method}

We work in the heavy-quark limit, so we use the non-relativistic expansion
of lattice QCD \cite{Caswell:1986ui,Eichten:1987xu,Lepage:1987gg}.
To evolve our quark propagators we use a time-step-symmetric form
for the transfer matrix \cite{Lepage:1992tx,Burch:2001tr}:
\BEA
\phi (\vec x,t+a) = \LP 1 - \frac{a{\cal H}_0}{2n}\RP_{t+a}^n U^\dagger_4(x)
\LP 1 - \frac{a{\cal H}_0}{2n}\RP_t^n \EL
\times (1 - \delta_{t',t} a\delta {\cal H})_t^{}\,\phi (\vec x,t) .
\EEA
The Hamiltonian applied at all time slices, ${\cal H}_0$, accounts for the
kinetic energy of the heavy quarks:
\BE
{\cal H}_0 = -\frac{\Delta^2}{2m_q} ,
\EE
where the $\Delta^2$ is the lattice covariant Laplacian. At one intermediate
time slice, $t'$, the lowest-order spin-dependent interaction,
\BE
\delta {\cal H} = -c_B^{}\frac{g}{2m_q}\vec\sigma\cdot\vec B ,
\EE
is applied, thereby allowing a spin flip of the quark (or antiquark) in
exchange for the emission or absorption of a gluonic excitation; i.e.,
configuration mixing.

The local value of the chromomagnetic field is calculated using the
clover formulation, averaging the fields generated from the four plaquettes
surrounding a lattice site:
\BEA
\Omega_{\mu\nu}(x) &=& \frac{1}{4}\left[ U_\mu(x)U_\nu(x+\hat{\mu})U^\dagger_\mu(x+\hat{\nu})U^\dagger_\nu(x)\right.\EL &\,& \hspace*{-1.5cm} \left.
+\,U_\nu(x)U^\dagger_\mu(x+\hat{\nu}-\hat{\mu})U^\dagger_\nu(x-\hat{\mu})U_\mu(x-\hat{\mu})\right.\EL &\,& \hspace*{-1.5cm} \left.
+\,U^\dagger_\mu(x-\hat{\mu})U^\dagger_\nu(x-\hat{\nu}-\hat{\mu})U_\mu(x-\hat{\nu}-\hat{\mu})U_\nu(x-\hat{\nu})\right.\EL &\,& \hspace*{-1.5cm} \left.
+\,U^\dagger_\nu(x-\hat{\nu})U_\mu(x-\hat{\nu})U_\nu(x-\hat{\nu}+\hat{\mu})U^\dagger_\mu(x) \right] ,
\EEA
\BE
{\cal F}_{\mu\nu}(x) = \frac{1}{2i}\left[ \Omega_{\mu\nu}(x) -
\Omega_{\mu\nu}^\dagger(x) \right]
- \frac{1}{3}{\rm Im}[{\rm Tr}\,\Omega_{\mu\nu}(x)] .
\EE
The chromomagnetic field arises from the spatial components,
\BE
{\cal F}_{jk}(x) = -\varepsilon_{jkl} g B_l(x) .
\EE
Tadpole improvement \cite{Lepage:1993xa} is also included, the factor $u_0$
being calculated via the average plaquette and applied to all the link
variables:
\BE
U_\mu(x) \rightarrow \frac{U_\mu(x)}{u_0} \hspace*{0.2cm};\hspace*{0.2cm}
{\cal F}_{\mu\nu}(x) \rightarrow \frac{{\cal F}_{\mu\nu}(x)}{u_0^4} .
\EE

Creating the heavy-quark propagators with this form for the evolution
operator, we then use appropriate operators at the source and sink time
slices to project out the desired meson states. The heavy-meson operators
we use, along with the corresponding quantum numbers, may be found in
Table 1 of Ref.~\cite{Burch:2001tr}.


Two different types of quark sources are used: a random wall (RW) and a
Coulomb-gauge-fixed wall (CW). The RW source is an incoherent collection of
point sources, the average meson propagator having contributions only from
where the quark and antiquark start at the same location. The CW source
provides a ``maximally smeared'' source, with contributions from spatially
separated quarks and antiquarks. In both cases, the sink end is simply a sum
over points where both the quark and antiquark coexist.

We fit the S-wave propagator with the following form:
\BE
C_s(t) = A_{1s} e^{-m_{1s}t} + A_{2s} e^{-m_{2s}t} ,
\label{swave_eq}
\EE
allowing a determination of the 2S-1S mass splitting. We also have P-wave
correlators, which we fit (at least for the CW source) with only a single
mass,
\BE
C_p(t) = A_{2p} e^{-m_{2p}t} .
\EE
We use the 2P-1S mass splitting to set the lattice scale. A single-mass fit
is also used for the hybrid correlators,
\BE
C_h(t) = A_{1h} e^{-m_{1h}t} .
\EE
All fits to the propagators use the full covariance matrix to account for
correlations among the different Euclidean times.

After application of the spin-dependent term in the Hamiltonian at $t'$
(or $t''$), a signal appears for the ``mixed'' correlator, hybrid
$\rightarrow$ S-wave (or vice versa). We fit these correlators in the region
$t > t'$ (or $t > t''$) with the forms:
\BE
C_{hs}^{(1)}(t',t) = A_{1hs}(t') e^{-m_{1s}(t-t')}
\EE
and
\BE
C_{sh}^{(1)}(t'',t) = A_{1sh}(t'') e^{-m_{1h}(t-t'')} .
\EE
Looking more closely at the amplitude for the first correlator (hybrid
$\rightarrow$ S-wave), we can reason that there should be factors
from the overlap of the source operator with the hybrid, the overlap of the
sink operator with the S-wave, the exponential decay of the hybrid state
before $t'$, and the matrix element with which we are interested:
\BE
A_{1hs}(t') = A_{1h}^{1/2}A_{1s}^{1/2}\LL 1S\left| a \delta{\cal H}
\right|1H\RR e^{-m_{1h}t'} .
\EE
Knowing the masses and amplitudes from the standard S-wave and hybrid
correlators, we solve for the off-diagonal matrix element of our
two-state Hamiltonian. This is repeated for larger values of $t'$ to find a
plateau in the final result, where we may be sure
that only the ground-state contribution from the source appears. A similar
procedure may be followed for the S-wave $\rightarrow$ hybrid correlator.

There is a complication, however, which arises for our CW-source
correlators: the operators at the source and sink ends are not the same.
At one end we have a Coulomb-gauge-fixed wall ($cw$) source, while at the
other end there is a point ($p$) sink. The amplitude for the S-wave
correlator therefore has the form
\BE
A_{1s} = A_{1s_{cw}}^{1/2} A_{1s_p}^{1/2} ,
\EE
while that for the hybrid is
\BE
A_{1h} = A_{1h_{cw}}^{1/2} A_{1h_p}^{1/2} .
\EE
From each of these products, the mixed correlators include only one of the
factors (rather than just the square root of each amplitude):
\BE
A_{1hs}(t') = A_{1s_p}^{1/2}A_{1h_{cw}}^{1/2}\LL 1S\left| a \delta{\cal H}
\right|1H\RR e^{-m_{1h}t'}
\label{hs_mixed_amp}
\EE
and
\BE
A_{1sh}(t'') = A_{1h_p}^{1/2}A_{1s_{cw}}^{1/2}\LL 1H\left| a \delta{\cal H}
\right|1S\RR e^{-m_{1s}t''} .
\label{sh_mixed_amp}
\EE
In order to get the appropriate cancellations of amplitudes, we thus need to
use a geometric mean:
\BE
\label{GEOM_MEAN}
\left| \LL 1S\left| a \delta{\cal H}\right|1H\RR \right| =
\sqrt{\frac{A_{1hs}(t')A_{1sh}(t'')}{A_{1s}A_{1h}} e^{m_{1h}t'} e^{m_{1s}t''}}
\EE
at large $t',t''$.

We average the correlators over sets of quenched lattices, generated
using a Symanzik 1-loop improved gauge action
\cite{Symanzik:1983dc,Luscher:1985zq,Lepage:1993xa,Alford:1995hw,
Bernard:1998mz}. The relevant parameters for these runs are listed in
Table \ref{NRQCD_RUNS}.

\begin{table}
\caption{
\label{NRQCD_RUNS}
Quenched NRQCD runs.
}
\begin{ruledtabular}
\begin{tabular}{ccccccc}
$\beta$ & $N_s^3\times N_t$ & $u_0$ & $am_q$ & $n$ & sources & \# configs. \\ \hline
7.75 & $16^3\times 32$ & 0.8800 & 3.2, 3.6 & 2 & RW,CW & 220 \\
8.00 & $20^3\times 64$ & 0.8879 & 2.5, 2.8 & 2 & RW,CW & 240 \\
$''$ & $''$ & $''$ & 0.7, 0.8 & 3 & CW & 170 \\
8.40 & $28^3\times 96$ & 0.89741 & 1.8, 2.0 & 2 & RW,CW & 76 \\
\end{tabular}
\end{ruledtabular}
\end{table}

To set the physical scale for the lattice spacing and determine the
physical quark mass, we use the same procedure as was used previously for
bottomonium spectroscopy with NRQCD \cite{Davies:1994mp}. For the lattice
spacings, we use bottomonium mass splittings, such as the spin-averaged
2P-1S mass difference ($M_{\bar\chi_b} - M_\Upsilon = 440$ MeV) and the
2S-1S difference for the $\Upsilon$ ($M_{\Upsilon'} - M_{\Upsilon} = 563$
MeV). We also create non-zero-momentum S-wave correlators and use the
resulting dispersion relations to determine the kinetic masses of the ground
states. Using the results from two quark masses, an interpolation (or
extrapolation) may be made to match the kinetic S-wave mass to that of the
$\Upsilon$ ($M_\Upsilon = 9.46$ GeV), thus arriving at a physical bottom
quark mass.

In an attempt to determine the radiative correction $c_B^{}$, we also
create correlators with the spin-dependent term applied at {\bf all}
intermediate time slices and with the values of $c_B^{} = 1$ and 2.
We then determine the resulting S-wave hyperfine splittings, which to
lowest order should be quadratic in this term:
$m_{1s}(1^{--})-m_{1s}(0^{-+}) \propto c_B^2$. There is, however, a danger
in assuming this to be the only (or the largest) contribution to the
hyperfine splitting: previous investigations of quarkonium spin-dependent
splittings in lattice NRQCD \cite{Trottier:1997ce,Stewart:2000ev} display
significant contributions from various $O(v^4)$ and $O(v^6)$ terms in the
velocity expansion (and poor convergence of this expansion for $c\bar c$
systems). Therefore, a value of $c_B^{}(a)$ determined in this way includes
not only the desired radiative correction, but also systematic effects due to
the neglect of other relativistic terms in the NRQCD expansion. To further
complicate the matter, the $0^{-+}$ bottomonium state ($\eta_b^{}$) has yet
to be observed experimentally.

\section{Results}

\begin{figure}
\resizebox{3.2in}{!}{\includegraphics{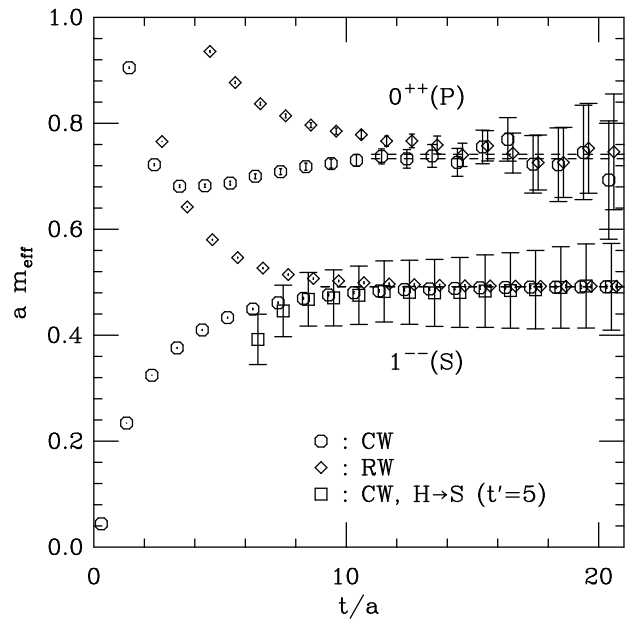}}
\caption{
\label{EFF_SP_QB800_M25}
Effective masses for the S-wave ($1^{--}$) and P-wave ($0^{++}$)
correlators from the $\beta=8.0$ lattices with $am_q=2.5$.
The dashed horizontal lines denote the masses ($am_{1s}^{}$, $am_{2p}^{}$)
from the (correlated) fits to the unperturbed CW-source correlators, along
with the corresponding fit ranges.
}
\end{figure}

\begin{figure}
\resizebox{3.2in}{!}{\includegraphics{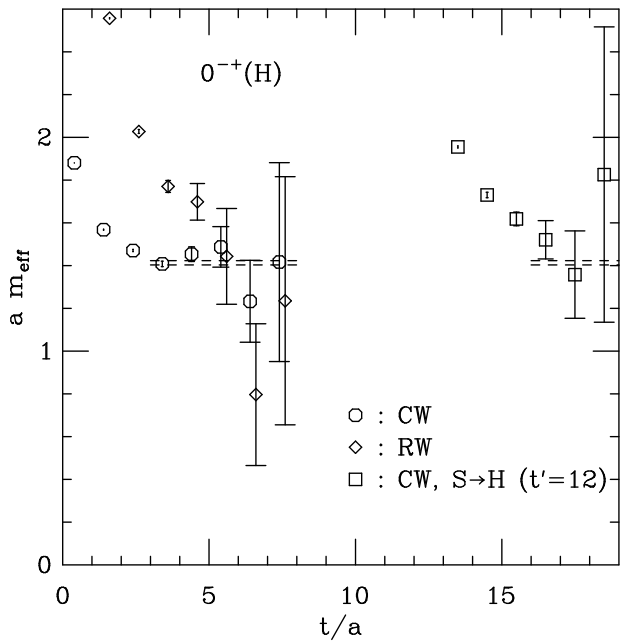}}
\caption{
\label{EFF_HYB_QB800_M25}
Effective masses for the $0^{-+}$ hybrid
correlators from the $\beta=8.0$ lattices with $am_q=2.5$.
The dashed horizontal lines denote the mass ($am_{1h}^{}$) from
the (correlated) fit to the unperturbed CW-source correlators, along with
the corresponding fit ranges.
}
\end{figure}

The Coulomb-gauge-fixed wall (CW) sources provide good overlap with the
desired meson states and reasonable fits are obtained earlier than for the
random wall (RW) sources. This can be easily seen in effective mass plots
of the P-wave (Fig.~\ref{EFF_SP_QB800_M25}) and hybrid correlators
(Fig.~\ref{EFF_HYB_QB800_M25}). The CW sources provide correlators which
approach plateaus earlier than their RW counterparts. The CW correlators
are less noisy as well; this is especially noticeable for the hybrids.
Both S-wave correlators take a substantially longer time to approach a
consistent plateau. These are reliably fit by the two-mass form found in
Eq.~(\ref{swave_eq}). Also included in the figures are examples of
the mixed correlators. In Fig.~\ref{EFF_SP_QB800_M25} the effective mass
for the H$\rightarrow$S correlator is shown for the region $t>t'=5$.
Although the mixed correlator is much more noisy, the agreement with the
mass from the unperturbed S-wave is good and we may fit this correlator
over a large range of $t$ to extract the mixed amplitude,
Eq.~(\ref{hs_mixed_amp}).
In Fig.~\ref{EFF_HYB_QB800_M25} we also show the effective mass for the
S$\rightarrow$H correlator for $t>t'=12$. Here we see the difficulty which
arises, namely the quality of signal, for the S$\rightarrow$H correlators
involving the point-like sinks. Nevertheless, for each value of $t'$,
we have at least a few significant points which give masses consistent with
those from the unperturbed CW-source hybrid correlator. We use these to
extract the corresponding mixed correlator amplitudes,
Eq.~(\ref{sh_mixed_amp}) in this case.
A complete list of the fits used for
our results, along with the amplitudes, masses, and $\chi^2$ values may be
found in Ref.~\cite{Burch:2003zz}.

As an aside, we point out the significant curvature in these plots for the
H$\rightarrow$S and S$\rightarrow$H correlators in the region $t'\approx t$,
even though the excited-state contributions from the sources should be
minimal: $e^{-m_{2h}^{}t'}$ and $e^{-m_{2s}^{}t'}$, respectively, become
small. These are most probably mixings with higher-mass configurations: e.g.,
$\LL 2S\left| a \delta{\cal H}\right|1H\RR$ and
$\LL 2H\left| a \delta{\cal H}\right|1S\RR$. In principle, the extraction
of these configuration-mixing amplitudes should also be possible, especially
with finer lattice spacing in the time direction (e.g., with anisotropic
lattices \cite{Juge:1999ie,Drummond:1999db}). For our purposes here, however,
we focus only upon the mixing between the lowest-lying configurations, 1S and
1H.


The results for the lattice spacings are presented in Table
\ref{LATTICE_SCALE}, along with other determinations from the static-quark
potential using a modified Sommer parameter $r_1$ \cite{Sommer:1994ce,
Bernard:2000gd} and the string tension $\kappa$. Looking at the NRQCD
results, we are encouraged by the consistency (within the errors of at least
one) of the 2P-1S and 2S-1S mass splittings. (We would like to point out,
however, that our heavy-quark Hamiltonian leaves out many terms in the
relativistic expansion and, for this reason, we do not claim to be presenting
a very precise determination of the bottomonium spectrum; this has been
studied by others using more elaborate forms of lattice NRQCD
\cite{Davies:1994mp,Davies:1998im,Davies:2003ik}.)
There are marked differences between the lattice spacings
determined via the bottomonium mass splittings and those from the static
quark potentials; in fact, the two static-quark determinations disagree.
The string-tension results consistently give the largest lattice spacings
($a$ in fm), while the $b\bar b$ mass splittings give the smallest. The
main reason for the difference of these two extremes appears to be the
quenched approximation. A comparison of the static-quark potentials for
quenched and dynamical configurations \cite{Bernard:2000gd} has shown that,
in the quenched case, the potentials do not display sufficient curvature:
the Coulomb-like potential well at short distances is not as deep as in the
dynamical case and the linear string-like region is steeper. This should
lead to an underestimation of the $b\bar b$ mass splittings in lattice
units since these systems are relatively small. This would seem to explain
the low values for the lattice spacings seen in Table \ref{LATTICE_SCALE}.
The string-tension results are thus expected to give the larger lattice
spacings for the quenched lattices. (A cursory study \cite{Burch:2003zz}
using the Born-Oppenheimer approximation and the quenched versus unquenched
potentials \cite{Bernard:2000gd} supports these claims.
Others \cite{Davies:1998im,Juge:1999ie} have also found, with better energy
resolution, discrepancies between 2S-1S and 2P-1S mass splittings on
quenched lattices, suggesting the same effect.)
In the spirit of presenting a self-consistent work, however, we use the
lattice spacings provided by the 2P-1S bottomonium mass splittings
\cite{endnote1}.

\begin{table}
\caption{
\label{LATTICE_SCALE}
Lattice spacing determinations with $c_B^{}=0$, CW source.
Physical scales used:
$r_1 = 0.344$ fm ; $\kappa^{1/2} = 440$ MeV ; 
$M_{\bar\chi_b} - M_\Upsilon = 440$ MeV ; 
$M_{\Upsilon'} - M_{\Upsilon} = 563$ MeV.
The * denote values used for $a$ throughout this work.
}
\begin{ruledtabular}
\begin{tabular}{ccll}
$\beta$ & $am_q$ & physical scale & $a^{-1}$ (MeV) \\ \hline
7.75 & $\infty$ & $r_1/a = 2.095(13)$ & 1200(7) \\
 & $\infty$ & $a^2\kappa = 0.1652(47)$ & 1082(15) \\
 & 3.2 & $a(M_{\bar\chi_b} - M_\Upsilon) = 0.328(13)$ & 1341(53)* \\
 & 3.6 & $a(M_{\bar\chi_b} - M_\Upsilon) = 0.324(12)$ & 1358(51) \\
 & 3.2 & $a(M_{\Upsilon'} - M_{\Upsilon}) = 0.487(60)$ & 1160(140) \\
 & 3.6 & $a(M_{\Upsilon'} - M_{\Upsilon}) = 0.434(34)$ & 1300(100) \\ \hline
8.00 & $\infty$ & $r_1/a = 2.6580(58)$ & 1522(3) \\
 & $\infty$ & $a^2\kappa = 0.09955(10)$ & 1394(7) \\
 & 2.5 & $a(M_{\bar\chi_b} - M_\Upsilon) = 0.2456(42)$ & 1792(31)* \\
 & 2.8 & $a(M_{\bar\chi_b} - M_\Upsilon) = 0.2431(40)$ & 1811(30) \\
 & 2.5 & $a(M_{\Upsilon'} - M_{\Upsilon}) = 0.3233(88)$ & 1742(48) \\
 & 2.8 & $a(M_{\Upsilon'} - M_{\Upsilon}) = 0.3131(73)$ & 1797(43) \\ \hline
8.40 & $\infty$ & $r_1/a = 3.7301(69)$ & 2136(4) \\
 & $\infty$ & $a^2\kappa = 0.04989(46)$ & 1970(9) \\
 & 1.8 & $a(M_{\bar\chi_b} - M_\Upsilon) = 0.1749(50)$ & 2516(71)* \\
 & 2.0 & $a(M_{\bar\chi_b} - M_\Upsilon) = 0.1724(47)$ & 2552(69) \\
 & 1.8 & $a(M_{\Upsilon'} - M_{\Upsilon}) = 0.235(16)$ & 2400(160) \\
 & 2.0 & $a(M_{\Upsilon'} - M_{\Upsilon}) = 0.229(14)$ & 2460(140) \\
\end{tabular}
\end{ruledtabular}
\end{table}

In Table \ref{1S_KIN_MASSES} we present our results for the 1S kinetic
masses determined from the dispersion relations:
\BE
E_{1s}(p) = \frac{p^2}{2M_{1S}^{kin}} + m_{1s} .
\EE
Each case requires an extrapolation in quark mass to reach the physical
value ($M_{1S}^{kin} = M_\Upsilon = 9.46$ GeV).

\begin{table}
\caption{
\label{1S_KIN_MASSES}
Kinetic masses for the 1S states and the resulting (lattice-regularized)
physical quark masses. The lattice scale is set using the
spin-averaged 2P-1S mass differences found for the $b\bar b$ systems.
}
\begin{ruledtabular}
\begin{tabular}{ccllll}
$\beta$ & $am_q$ & $aM_{1S}^{kin}$ & $M_{1S}^{kin}$ (GeV) & $am_b$ & $m_b$ (GeV) \\ \hline
7.75 & 3.2 & 7.38(54) & 9.90(72) & 3.09(21) & 4.17(28) \\
 & 3.6 & 8.44(59) & 11.46(80) &  &  \\ \hline
8.0 & 2.5 & 5.509(77) & 9.87(14) & 2.41(4) & 4.34(7) \\
 & 2.8 & 6.178(90) & 11.19(16) &  &  \\ \hline
8.4 & 1.8 & 4.00(16) & 10.06(40) & 1.70(7) & 4.31(18) \\
 & 2.0 & 4.42(17) & 11.28(43) &  &  \\ \hline
 & & & & $am_c$ & $m_c$ (GeV) \\ \hline
8.0 & 0.7 & 1.752(23) & 3.154(41) & 0.677(12) & 1.219(22) \\
 & 0.8 & 1.964(26) & 3.535(47) &  &  \\
\end{tabular}
\end{ruledtabular}
\end{table}

\begin{figure}
\resizebox{3.2in}{!}{\includegraphics{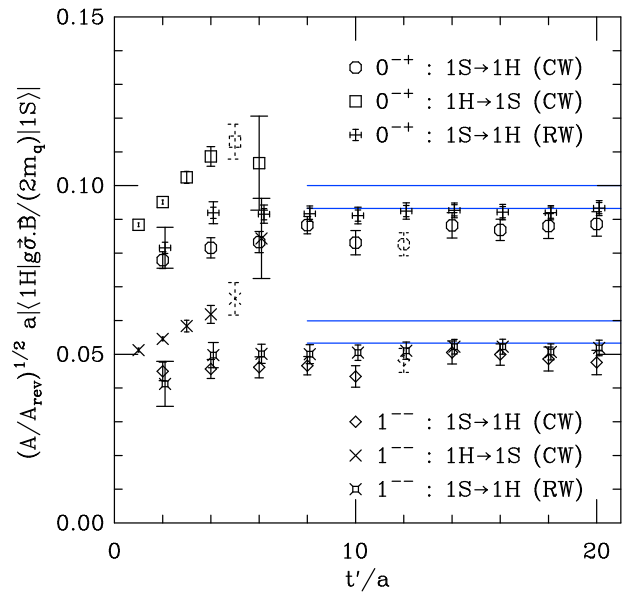}}
\caption{
\label{SIGMAB_CGF_NGF_QB800}
Comparison of the results obtained for the mixing matrix element via the
CW source and the RW source with the hybrid mass fixed to that found with
the CW hybrid source. The dotted symbols mark the ones used in the
geometric mean to get the CW-source results, the $1\sigma$ ranges of which
are denoted by the horizontal lines.
}
\end{figure}

\begin{figure}
\resizebox{3.2in}{!}{\includegraphics{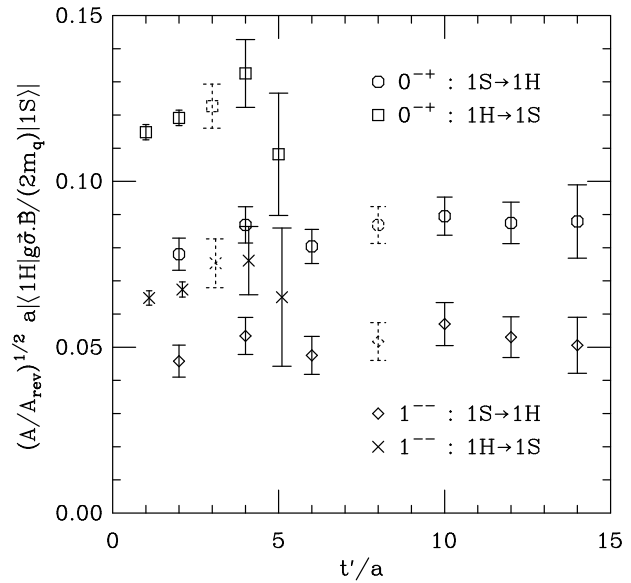}}
\caption{
\label{SIGMAB_CGF_QB775}
The mixing matrix element {\it vs} the time of application of the
interaction term. CW source, $\beta=7.75$, $am_q=3.2$.
The dotted symbols mark the ones used in the geometric mean to get the
CW-source results.
}
\end{figure}

\begin{figure}
\resizebox{3.2in}{!}{\includegraphics{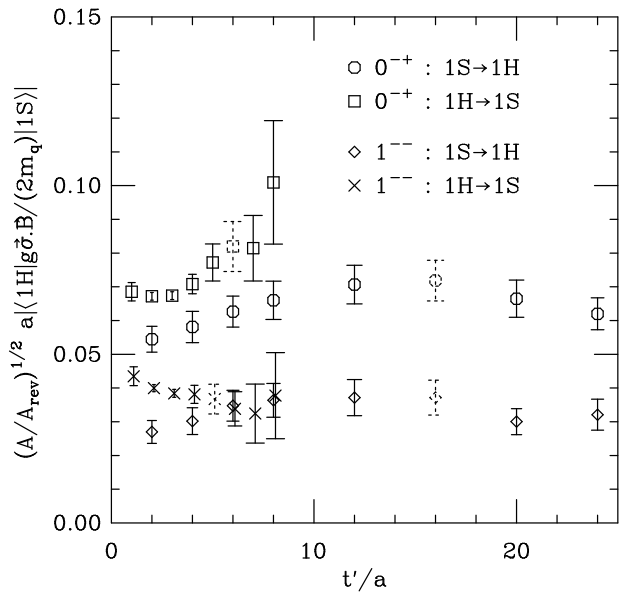}}
\caption{
\label{SIGMAB_CGF_QB840}
The mixing matrix element {\it vs} the time of application of the
interaction term. CW source, $\beta=8.4$, $am_q=1.8$.
The dotted symbols mark the ones used in the geometric mean to get the
CW-source results.
}
\end{figure}

\begin{figure}
\resizebox{3.2in}{!}{\includegraphics{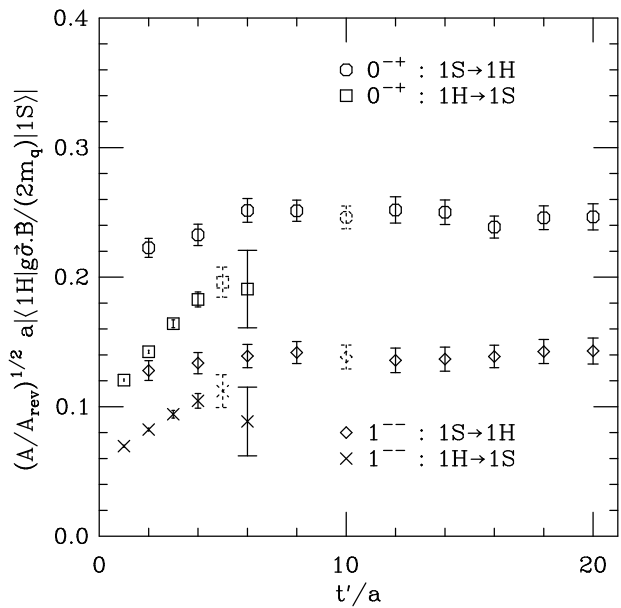}}
\caption{
\label{SIGMAB_CGF_QB800_CC}
The mixing matrix element {\it vs} the time of application of the
interaction term. CW source, $\beta=8.0$, $am_q=0.7$.
The dotted symbols mark the ones used in the geometric mean to get the
CW-source results.
}
\end{figure}

As described in the previous section, we use the amplitudes from the mixed
correlators, along with the masses and amplitudes from the ``unmixed''
ones, to determine the off-diagonal element of our two-state Hamiltonian.
Figures \ref{SIGMAB_CGF_NGF_QB800}$-$\ref{SIGMAB_CGF_QB800_CC} show the
results for these matrix elements as functions
of the time, $t'$, at which the spin-dependent term is applied to the
heavy-quark propagators. The CW-source, H$\rightarrow$S results each have the
remaining factor of
\BE
\frac{A_{1s_p}^{1/2}A_{1h_{cw}}^{1/2}}{A_{1s}^{1/2}A_{1h}^{1/2}}
= \frac{A_{1s_p}^{1/4}A_{1h_{cw}}^{1/4}}{A_{1s_{cw}}^{1/4}A_{1h_p}^{1/4}}
= \LP\frac{A}{A_{rev}}\RP^{1/2}
\EE
(the reciprocal for S$\rightarrow$H) which requires the use of a
geometric mean (see Eq. \ref{GEOM_MEAN}) to ensure its removal.
The results chosen for the final geometric means are displayed with dotted
symbols. All errors result from a single-elimination jackknife routine.

Figure \ref{SIGMAB_CGF_NGF_QB800} displays results for the
configuration-mixing matrix element from one set of lattices ($\beta=8.0$)
and one value of the quark mass ($am_q=2.5$). Results are shown using both
the CW and RW sources. The hybrid correlators for the latter, however, have
their masses fixed to the values found from the CW-source hybrid correlators
since these provide more reliable mass plateaus. The two horizontal lines
show the $\pm 1\sigma$ limits for the geometric mean of the CW-source results.
The hybrid$\rightarrow$S-wave results do not display very convincing
plateaus, especially for the $1^{--}$ channel. However, within the errors,
the final results are consistent with the plateaus from the RW source.
These results appear to be about 30\% lower than previous determinations
using only the RW sources \cite{Burch:2001tr,Burch:2001nk}, thereby stressing
the need for reliable hybrid mass determinations, which the CW-source
correlators more readily provide (see Fig.~\ref{EFF_HYB_QB800_M25}).

Figures \ref{SIGMAB_CGF_QB775} and \ref{SIGMAB_CGF_QB840} display similar
plots for the $\beta=7.75$ and 8.4 lattices, respectively. Here we see more
convincing plateaus for the matrix element from the hybrid$\rightarrow$S-wave
correlators.

In Fig.~\ref{SIGMAB_CGF_QB800_CC} we show the results for one of the lighter
masses (around $m_c^{}$) for the $\beta=8.0$ lattices.
A quick look at the result for the $1^{--}$ channel,
however, shows that the non-relativistic approximation (at least at the
level of simplicity in our heavy-quark Hamiltonian) may not be such a great
idea:
\BE
\frac{\langle 1H|\delta{\cal H}|1S\rangle}{2m_q} \approx \frac{0.14}{1.4}
\approx 0.1 \sim v^4 .
\EE
The mass required at this lattice spacing to reach the charm quark is also
problematic for the expansion ($am_c < 1$). In spite of these problems we
carry on and present our results for charmonium, encouraging the reader not
to forget the large systematic effects we introduce by neglecting other
terms in our heavy-quark Hamiltonian and by simulating at such a small quark
mass.

All of the CW-source results for the configuration-mixing matrix elements,
along with the corresponding mixing angles, are shown in Tables
\ref{SIGMA_B_SIN_THETA_1} and \ref{SIGMA_B_SIN_THETA_0}. The last column
displays the final results after an extrapolation (linear) in quark mass
to the bottom (or charm) mass determined previously (see Table
\ref{1S_KIN_MASSES}). The hybrid configuration content of the $0^{-+}$ ground
state is enhanced by the expected factor of $\sqrt{3}$ (due to spin
statistics for single-gluon emission/absorption
\cite{Barnes:1979hg,Barnes:1983tx}) relative to that in the
$1^{--}$ channel. Up to this point we have ignored the radiative
correction ($c_B^{}=1$) and we point out the need for this factor,
$c_B^{}(a(\beta))$, in the final column. To be precise, this factor should
be included in the matrix element, but for small mixing angles, this
correction appears as a multiplicative factor in the final result:
$\sin\theta\approx\langle1H|\delta{\cal H}|1S\rangle/(m_{1h}-m_{1s})$.

\begin{table}
\caption{
\label{SIGMA_B_SIN_THETA_1}
Results for the $1^{--}$ S-wave/hybrid configuration mixing (CW source).
}
\begin{ruledtabular}
\begin{tabular}{ccccc}
$\beta$ & $am_q$ & $\left|\LL 1H\left| a \delta{\cal H} \right| 1S\RR\right|$ & $|\sin(\theta)|$ & $|\langle 1H|\Upsilon\rangle|$ \\ \hline
7.75 & 3.2 & 0.0624(61) & 0.0566(27) & 0.058(3)$c_B^{}(\beta)$ \\
 & 3.6 & 0.0582(61) & 0.0518(27) &  \\ \hline
8.0 & 2.5 & 0.0566(33) & 0.0618(28) & 0.063(3)$c_B^{}(\beta)$ \\
 & 2.8 & 0.0522(31) & 0.0564(27) &  \\ \hline
8.4 & 1.8 & 0.0369(44) & 0.0618(36) & 0.064(4)$c_B^{}(\beta)$ \\
 & 2.0 & 0.0352(44) & 0.0578(36) &  \\ \hline
 & & & & $|\langle 1H|J/\psi\rangle|$ \\ \hline
8.0 & 0.7 & 0.1349(84) & 0.1483(81) & 0.150(8)$c_B^{}(\beta)$ \\
 & 0.8 & 0.1259(74) & 0.1393(70) &  \\
\end{tabular}
\end{ruledtabular}
\end{table}

\begin{table}
\caption{
\label{SIGMA_B_SIN_THETA_0}
Results for the $0^{-+}$ S-wave/hybrid configuration mixing (CW source).
}
\begin{ruledtabular}
\begin{tabular}{ccccc}
$\beta$ & $am_q$ & $\left|\LL 1H\left| a \delta{\cal H} \right| 1S\RR\right|$ & $|\sin(\theta)|$ & $|\langle 1H|\eta_b^{}\rangle|$ \\ \hline
7.75 & 3.2 & 0.1032(58) & 0.0945(26) & 0.097(3)$c_B^{}(\beta)$ \\
 & 3.6 & 0.0954(57) & 0.0862(26) &  \\ \hline
8.0 & 2.5 & 0.0966(34) & 0.1031(29) & 0.106(3)$c_B^{}(\beta)$ \\
 & 2.8 & 0.0887(32) & 0.0939(27) &  \\ \hline
8.4 & 1.8 & 0.0767(60) & 0.1156(52) & 0.120(5)$c_B^{}(\beta)$ \\
 & 2.0 & 0.0723(61) & 0.1074(53) &  \\ \hline
 & & & & $|\langle 1H|\eta_c^{}\rangle|$ \\ \hline
8.0 & 0.7 & 0.2327(77) & 0.2403(62) & 0.243(6)$c_B^{}(\beta)$ \\
 & 0.8 & 0.2172(69) & 0.2270(56) &  \\
\end{tabular}
\end{ruledtabular}
\end{table}

Meson correlators with the $\vec\sigma\cdot\vec B$ term applied at all
intermediate time slices were also created in order to study the S-wave
hyperfine splittings and to attempt to determine $c_B^{}(\beta)$.
For these, the values of $c_B^{}=1$ and 2 were intended; however, a sign
error was discovered (much too late) in our field-strength routine which
forces us to work with $c_B^{}=-1$ and $-2$. (This has no effect upon our
configuration-mixing results up to this point since they result from a
single application of the spin-dependent term.) The resulting hyperfine
splittings for the S-wave states may be found in Table \ref{HYP_SPLIT}.

\begin{table}
\caption{
\label{HYP_SPLIT}
1S hyperfine splittings.
}
\begin{ruledtabular}
\begin{tabular}{ccccc}
$\beta$ & $am_q$ & $c_B^2$ & $M_\Upsilon - M_{\eta_b}$ (MeV) & $c_B^2(\beta)$ \\ \hline
7.75 & 3.2 & 1 & 16.7(1.3) & $(M_\Upsilon - M_{\eta_b})_{\rm exp}/(17\,{\rm MeV})$ \\
 & & 4 & 63.6(5.1) & \\ \hline
8.0 & 2.5 & 1 & 21.90(55) & $(M_\Upsilon - M_{\eta_b})_{\rm exp}/(22\,{\rm MeV})$ \\ 
 & & 4 & 81.6(2.6) & \\ \hline
8.4 & 1.8 & 1 & 25.2(1.1) & $(M_\Upsilon - M_{\eta_b})_{\rm exp}/(25\,{\rm MeV})$ \\
 & & 4 & 93.5(4.9) & \\ \hline
 & & & $M_{J/\psi} - M_{\eta_c}$ (MeV) \\ \hline
8.0 & 0.7 & 1 & 68.0(1.8) & $\approx 1.2-1.7$ \\
 & & 4 & 213.6(7.1) & \\
\end{tabular}
\end{ruledtabular}
\end{table}

Whereas our bottomonium
systems seem to follow the $c_B^2$ proportionality and our charmonium systems
do not, there is no experimentally measured mass for the $0^{-+}$ $b\bar b$
meson ($\eta_b^{}$) while the corresponding $c\bar c$ meson ($\eta_c^{}$) has
been observed. So we report our findings for $c_B^2(a(\beta))$ from the
bottomonium systems in terms of the $M_\Upsilon-M_{\eta_b}$ mass splitting.
For the charmonium results, we report two values for $c_B^2(\beta)$.
The larger value is the result of a linear extrapolation from 0 through
the $c_B^2=1$ point to $M_{J/\psi}-M_{\eta_c}=117$ MeV at $c_B^2=c_B^2(\beta)$.
If we believe the physical quark mass shifts with the inclusion of the
spin-dependent term and that this is the dominant effect in explaining why
our $c\bar c$ hyperfine splitting is not $\propto c_B^2$, then this is the
result we would trust more. If, however, we believe there is a $c_B^3$
correction present and that this is the dominant effect, we should trust
the lower value, which results from using an interpolation of the form
$a_2c_B^2 + a_3c_B^3$ through the $c_B^{}=-1$ and $-2$ points and
extrapolating to 117 MeV in the $c_B^{} > 0$ region.
If we use these results with that from $b\bar b$ at $\beta=8.0$, we find
$M_\Upsilon-M_{\eta_b} \approx 26-38$ MeV and
$c_B^{}(\beta=7.75) \approx 1.24-1.50$,
$c_B^{}(\beta=8.0) \approx 1.09-1.31$,
$c_B^{}(\beta=8.4) \approx 1.02-1.23$.
Any such determination of the radiative corrections, however, is further
complicated by the fact that it incorporates systematic effects due to the
termination of the NRQCD expansion \cite{Trottier:1997ce,Stewart:2000ev}.

Given our previous worries about our $c\bar c$ results ($v^4 \sim 0.1$;
$am_c < 1$), we may also consider the possible range of $c_B^{}$ values
using potential model values for the $b\bar b$ S-wave hyperfine splittings
\cite{Chen:1996bc,Yndurain:2000yq}: $\sim 30-60$ MeV. Assuming the
$\vec\sigma\cdot\vec B$ term to dominate this splitting results in
$c_B^{}(\beta=7.75) \sim 1.33-1.88$,
$c_B^{}(\beta=8.0) \sim 1.17-1.65$,
$c_B^{}(\beta=8.4) \sim 1.10-1.55$; or roughly
\BEA
|\langle 1H|\Upsilon\rangle| \sim 0.076 - 0.11
\hspace*{0.2cm};\hspace*{0.2cm}
|\langle 1H|\eta_b^{}\rangle| \sim 0.13 - 0.19 \nonumber
\EEA
and
\BEA
|\langle 1H|J/\psi\rangle| \sim 0.18 - 0.25
\hspace*{0.2cm};\hspace*{0.2cm}
|\langle 1H|\eta_c^{}\rangle| \sim 0.29 - 0.4 . \nonumber
\EEA
Each of these ranges lies between two corresponding values (with and without
the color Coulomb interaction) determined within the MIT bag model
\cite{Barnes:1979hg}. The upper limit of the $J/\psi$ result is consistent
with that from an adiabatic potential model \cite{Gerasimov:1998tm}.

Again, however, the above determinations of $c_B^{}(\beta)$ include
systematic effects due to the neglect of other terms in the velocity
expansion which also contribute to the hyperfine splittings
\cite{Trottier:1997ce,Stewart:2000ev}.
We are therefore left without precise values for the radiative corrections
and may only state that if the $\vec\sigma\cdot\vec B$ term does in fact
dominate the S-wave hyperfine splitting, then these corrections appear to
enhance the configuration mixing further.

\begin{figure}
\resizebox{3.2in}{!}{\includegraphics{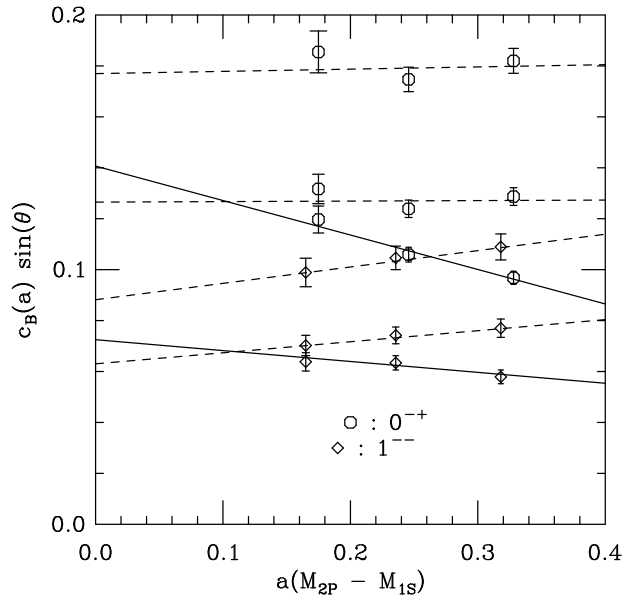}}
\caption{
\label{SIN_S_VS_A}
Hybrid/S-wave configuration mixing angle {\it vs} lattice spacing.
For each channel, the three lowest points (solid fit line) represent the
(tadpole-improved) tree-level results: $c_B^{}=1$.
The middle three points result from setting the radiative corrections with
the S-wave hyperfine splitting held constant at 30 MeV.
The uppermost points result from the same splitting set to 60 MeV.
The $1^{--}$ points are shifted slightly to the left for clarity.
}
\end{figure}

In Fig.~\ref{SIN_S_VS_A} we plot our results for the bottomonium
configuration-mixing angle as a function of the lattice spacing.
The plot includes the tree-level results, along with two others altered
using radiative corrections given by assumed values for the S-wave hyperfine
splitting due to the $\vec\sigma\cdot\vec B$ term.
Some positive features may be noted in this plot.
First, the choice of the ``physical'' value for the hyperfine splitting does
not seem to have a large effect upon the scaling displayed by the points
(however, the values at any one lattice spacing are greatly affected).
Also, at least for the $0^{-+}$ channel, the fixing of the hyperfine
splittings appears to slightly improve the scaling; the tree-level and
radiatively-corrected values for the $1^{--}$ appear to scale equally well.

For the bottomonium systems on our lattices, where we have
$a \gtrsim 1/m_b^{}$, discretization effects are expected to be about the
same order as those due to the relativistic corrections we neglect.
In Ref.~\cite{Lepage:1992tx}, a prescription for removing such systematics
by improving the heavy-quark action has been presented.
Such improvements are not included in this current work and therefore
discretization effects may also be partly responsible for the trends
seen in the tree-level results of Fig.~\ref{SIN_S_VS_A}
(for the $0^{-+}$ channel, $\sim$20\% rise in the central value by cutting
the lattice spacing in half; $\sim$10\% for the $1^{--}$).

\section{Conclusions}

We have determined the relative contribution of the lowest hybrid
configuration to the wavefunctions of heavy S-wave mesons. Our approach
utilizes the non-relativistic approximation to lattice QCD, applicable for
heavy quarkonia where there is a clear separation of radial, orbital, and
gluonic excitations and the mixing among the corresponding configurations
is small.

Tadpole-improved, tree-level results ($c_B^{}=1$ in Tables
\ref{SIGMA_B_SIN_THETA_1} and \ref{SIGMA_B_SIN_THETA_0}) display hybrid
configuration admixtures at about $(0.063)^2 \approx 0.4$\% probability
within the $\Upsilon$ and about $(0.15)^2 \approx 2.3$\% for $J/\psi$.
The corresponding results for the pseudoscalar channels are enhanced by
$\approx 3$ due to spin statistics \cite{Barnes:1979hg,Barnes:1983tx}.

Although the radiative corrections for the spin-dependent term in our
heavy-quark Hamiltonian elude evaluation, these factors may provide
further enhancement of the configuration mixing.

Quenching appears to significantly affect the lattice spacing results
(see Table \ref{LATTICE_SCALE} and Ref.~\cite{Bernard:2000gd}), decreasing
those determined via bottomonium mass splittings (by perhaps 10\%, see
Ref.~\cite{Burch:2003zz}). This, in turn, affects the quark mass
extrapolation and the determination of the radiative correction, but the
direct effect upon the configuration mixings is not clear. This may only be
answered by repeating the analysis on lattices with dynamical quarks.

Inclusion of higher-order terms within the NRQCD Hamiltonian has been shown
to alter bottomonium hyperfine splittings significantly (by as much as 15\%
in one case \cite{Trottier:1997ce}). In principle, such terms can cause the
same configuration mixings as those determined here through only the
$\vec\sigma\cdot\vec B$ term and may be needed to achieve a more accurate
measure of the hybrid configuration content within the $\Upsilon$ and
$\eta_b^{}$. The situation for our charmonium results is more dire. For these
systems, the heavy-quark expansion can hardly be trusted on our lattices
($am_c < 1$) and higher-order corrections may contribute as much or more to
the mixing.

These results have implications for theoretical determinations of certain
quarkonium quantities. For example, since the $J/\psi \rightarrow e^+e^-$
decay should not occur directly from the hybrid configuration
\cite{Barnes:2003zz}, our results suggest about a
$(0.15)^2 \approx 2.3(2)$\% suppression of the
$\Gamma(J/\psi \rightarrow e^+e^-)$ partial width determined
via potential models; about a $(0.063)^2 \approx 0.40(4)$\%
suppression for $\Gamma(\Upsilon \rightarrow e^+e^-)$. These effects appear
to be rather small (the corresponding experimental errors lie at about 7\%
and 4\% for $J/\psi$ and $\Upsilon$, respectively \cite{Hagiwara:2002pw})
and this analysis lends support to their neglect thus far, at least for
the $\Upsilon$, where our lattice heavy-quark expansion is more trustworthy.

The other side to this issue is that of the vector hybrid states containing
some admixture of the $q\bar q$ vector configurations and therefore coupling
to $e^+e^-$. The above (tree-level) results would suggest the appearance of
a $c\bar cg$ hybrid resonance with a partial width of
\BEA
\Gamma(c\bar cg \rightarrow e^+e^-) &\approx& 0.023(2) \times
\Gamma(J/\psi \rightarrow e^+e^-) \EL
&\approx& 0.12(1) \, {\rm keV} \nonumber
\EEA
somewhere around $E_{cm} = M_{c\bar cg} \approx 4.4 - 4.7$ GeV (the lower
estimate arises from using the $r_1^{}$-determined value of the lattice
spacing) and a $b\bar bg$ hybrid resonance with
\BEA
\Gamma(b\bar bg \rightarrow e^+e^-) &\approx& 0.0040(4) \times
\Gamma(\Upsilon \rightarrow e^+e^-) \EL
&\approx& 0.0053(5) \, {\rm keV} \nonumber
\EEA
around $M_{b\bar bg} \approx 10.8 - 11.1$ GeV. Any such $b\bar bg$-dominated
resonance should thus be extremely difficult to see in $e^+e^-$ $R$-scans
given this very small partial width and (most likely) much larger overall
width. The $c\bar cg$-dominated state gives a partial width of the same order
of that seen in a resonance within the same mass range:
$\psi(4415)$ with $\Gamma(e^+e^-) = 0.47(10)$ keV
\cite{Hagiwara:2002pw,Siegrist:1976br,Brandelik:1978ei};
$\psi(4430)$ with $\Gamma(e^+e^-) = 0.390(74)$ keV from
Refs.~\cite{Bai:2001ct,Eidemuller:2002ru}.
However, this hardly qualifies as evidence of a hybrid-dominated
structure for such a resonance. Our charmonium results are plagued with
systematic errors which are not easily quantified (given the charm quark
mass and lattice spacings we use, the mixing through other terms in the
expansion of the heavy-quark Hamiltonian may be even larger). Besides, any
such determination would not only require a study of the possible decays from
the hybrid state (determining the total width), but would also need to
consider mixings of the $c\bar cg$ configuration with other $c\bar c$
configurations which lie closer in mass.
The use of lattices with a finer temporal resolution may be helpful in
resolving multiple states in future simulations.

\begin{acknowledgments}
This work was supported by the U.S. Department of Energy under contract
DE FG03 95ER 40906. Computations were performed on Blue Horizon at the
San Diego Supercomputing Center and the Nirvana cluster at Los Alamos
National Laboratory. We would like to thank Ted Barnes for helpful
suggestions.
\end{acknowledgments}

\bibliography{heavy_config_mix}

\begin{thebibliography}{37}
\bibitem[{end()}]{endnote1}
\bibinfo{note}{The current SP splitting for the $\beta=8.0$ lattices is about
  10\% lower than that reported previously in Ref.~[7]. Given our current
  results from higher statistics, two source operators and consistent masses
  from the two (see Fig.~1 or Fig.~5.2 of Ref.~[9]), we feel we have
  effectively removed such systematics.}
\end{thebibliography}

\end{document}